\documentstyle[12pt,cite]{article}
  \def\half{{1 \over 2}}
  \def\i3{{1 \over 3}}
  \def\sqr2{\sqrt{2}}
  \def\isqr2{{1 \over \sqrt{2}}}

  \def\tO{\tilde{\Omega}}

  \def\to{\tilde{\omega}}

\begin{document}

\thispagestyle{empty}
\vspace*{-15mm}
\baselineskip 10pt
\begin{flushright}
\begin{tabular}{l}
{\bf CERN-TH/98-239}\\
{\bf hep-ph/9807449}
\end{tabular}
\end{flushright}

\vspace{2cm}
\begin{Large}
\begin{center}
Single and Double Universal Seesaw Mechanisms\\ with\\ Universal Strength
for Yukawa Couplings
\end{center}
\end{Large}

\vskip 0.5cm
\begin{large}
\begin{center}
I. S. Sogami${}^{a,b}$,
H. Tanaka${}^{a}$
and T. Shinohara${}^{a}$
\\
\vskip 0.3cm
{\it 
${}^a$\ Department of Physics, Kyoto Sangyo University, Kyoto 603-8555, Japan\\
${}^b$\ Theoretical Physics Division, CERN, CH-1211 Geneva 23, Switzerland}
\end{center}
\end{large}

\vskip 1.0cm
\begin{abstract}
\noindent
Single and double universal seesaw mechanisms and the hypothesis of
{\it universal strength for Yukawa couplings} are applied to
formulate a unified theory of fermion mass spectrum in a model based on
an extended Pati-Salam symmetry. Five kinds of Higgs fields are postulated
to mediate scalar interactions among electroweak doublets of light fermions
and electroweak singlets of heavy exotic fermions with relative
Yukawa coupling constants of exponential form.
At the first-order seesaw approximation, quasi-democratic mass matrices with
equal diagonal elements are derived for all charged fermion sectors and a
diagonal mass matrix is obtained for the neutrino sector under an additional
ansatz. Assuming the vacuum neutrino oscillation, the problems of solar
and atmospheric neutrino anomalies are investigated.
\end{abstract}

\vspace{20mm}
\noindent
\begin{tabular}{l}
{\bf CERN-TH/98-239}\\
July 1998
\end{tabular}

\newpage

\baselineskip 18pt

\section{Introduction}

The hypothesis of the universal seesaw mechanism
(USM)~\cite{useesaw,soga1,soga2,koide} was invented to explain the smallness
of the charged-fermion masses relative to the electroweak scale by
postulating the existence of exotic fermions belonging to electroweak singlets.
Although the large observed value of the top-quark mass~\cite{topmass} seems
to reduce its original merit as far as the up-quark sector is concerned, it is
effective for the down-quark and charged-lepton sectors. Furthermore,
the double universal seesaw mechanism~\cite{soga1,soga2} is very
promising to explain the origin of the tiny masses of neutrinos,
for which experimental evidence has been strongly sustained by
the recent observation of an atmospheric neutrino anomaly at
the Super-Kamiokande~\cite{superk}.

We have developed a unified theory of quark-lepton mass spectrum with the single
and double universal seesaw mechanisms (SDUSM)~\cite{soga1,soga2} in the model
based on the extended Pati-Salam gauge group~\cite{patisalam},
\begin{equation}
  G \equiv SU(4)_{cL}\times SU(4)_{cR}\times SU(2)_L\times SU(2)_R\times U(1)_X,
  \label{patisalam}
\end{equation}
where a group $U(1)_X$ is generated by a new charge $X$. In the theory,
the $U(1)$ chiral charges~\cite{chiral} were assigned to the fermion
and Higgs fields to distinguish generations and to restrict patterns
of the Yukawa couplings. Mass matrices of the same extended Fritzsch
type~\cite{fritzsch} were obtained for all four fermion sectors and
the problem of solar neutrino deficit was analysed by assuming
the Mikheyev-Smirnov-Wolfenstein mechanism~\cite{msw} and the vacuum
neutrino oscillation~\cite{vacoscillation}. \par

The purpose of this article is to formulate a new unified theory of quark-lepton
mass spectrum by combining the SDUSM with the hypothesis of {\it universal
strength for Yukawa couplings} (USY)~\cite{branco,usyukawa} in the model
based on the same gauge group $G$ in Eq.\,(\ref{patisalam}). We obtain
quasi-democratic mass matrices~\cite{democratic} of specific forms with equal
diagonal elements for the charged-fermion sectors.
The hypothesis of the USY is proved to work so favourably on the USM in our
formalism that it is applicable also to the up-quark sector~\cite{soga3} .
With an additional ansatz, a diagonal mass matrix is obtained for the neutrino
sector. We examine the problems of the solar and atmospheric neutrino anomalies
by assuming the vacuum neutrino oscillations. \par

In the SDUSM,  two kinds of exotic fermion multiplets belonging to electroweak
singlets, i.e. colour quartets $F_{ih} (F = U, D)$ and colour singlets
$N'_{ih}$, are postulated to exist for each generation of colour quartets
$\psi_{ih}$  belonging to electroweak doublets, where
$i\ (i = 1,\,2,\,3)$ and $h\ (h = L, R)$ stand for, respectively, the generation
and the chirality. Couplings between the electroweak doublets $\psi_{iL}$ and
$\psi_{iR}$ are forbidden by the structure of the fundamental gauge group $G$.
We naturally assume that the purely-neutral fields $N'_{ih}$ have intrinsic
bare Dirac masses. Therefore, the {\it Yukawa couplings} can exist between
$\psi_{ih}$ and $F_{ih}$ of colour quartets, between $F_{ih}$ and $N'_{ih}$ of
electroweak singlets, and between $F_{ih}$ themselves. Extending the original
hypothesis of the USY~\cite{branco}, we require that all three types of these
Yukawa couplings depend on generations by their phase factors.\par

Note that the hypothesis of the USY in the standard model~\cite{branco} is
specified in the gauge-interaction eigenmodes. In our model with the SDUSM,
however, there is a large freedom for fermion bases, which allows equivalent
eigenmodes for the gauge interaction. For example, arbitrary unitary
transformations can be applied to the neutral exotic fields $N'_{ih}$
without changing the gauge interaction. Therefore, in order to state the USY
strictly in our model, it is necessary to select a certain preferred base
for the gauge interaction eigenmodes (see {\S}2). \par

\section{Model}

Fundamental fermions are classified with respect to the basic group $G$.
Dominant components of ordinary quarks and leptons belonging to
the $i$-th generation are described, generically, by the chiral fields
$\psi_{iL}$ and $\psi_{iR}$ with the following transformation properties:
 \begin{equation}
   \psi_{iL} \sim (4, 1, 2, 1; 0), \quad \psi_{iR} \sim (1, 4, 1, 2; 0).
 \end{equation}
In order to implement the USM, it is necessary to introduce the electroweak
singlets of chiral fermion fields
 \begin{equation}
  \begin{array}{ll}
   U_{iL} \sim (1, 4, 1, 1; 1) ,\ &
   U_{iR} \sim (4, 1, 1, 1; 1) ,\\
   D_{iL} \sim (1, 4, 1, 1; -1),\ &
   D_{iR} \sim (4, 1, 1, 1; -1) \\
  \end{array}
 \end{equation}
as the seesaw partners for each generation. Note here that the colour gauge
fields of $SU(4)_L$ and $SU(4)_R$ groups interact with the fermion multiplets
$(\psi_{iL}, U_{iR}, D_{iR})$ and $(\psi_{iR}, U_{iL}, D_{iL})$, respectively.
With these specific choice of the fermion multiplets, the cancellation of
triangular anomalies~\cite{anomaly1} is ensured even when the exact left-right
symmetry in the gauge interaction is broken~\cite{anomaly2}. For the USM to be
induced doubly in the neutral fermion sector, electroweak singlets of
colourless fermions described by the chiral fields
 \begin{equation}
   N_{iL}' \sim (1, 1, 1, 1; 0) , \quad
   N_{iR}' \sim (1, 1, 1, 1; 0)
 \end{equation}
are postulated to exist per generation.\par

Symmetry breakings from $SU(4)_{cL}\times SU(4)_{cR}$ down to $SU(3)_c$ through
$SU(3)_{cL}\times SU(3)_{cR}$ are built in by colour quartets and a bi-quartet of
Higgs fields
 \begin{equation}
   \chi_{cL} \sim (4, 1, 1, 1; 1), \quad \chi_{cR} \sim (1, 4, 1, 1; 1),
   \quad \Phi \sim (4, \overline{4}, 1, 1; 0),
 \end{equation}
which develop vacuum expectation values
 \begin{equation}
   \langle \chi_{cL} \rangle = \left(
                \matrix{0 \cr 0 \cr 0 \cr v_L \cr} \right) , \quad
   \langle \chi_{cR} \rangle = \left(
                \matrix{0 \cr 0 \cr 0 \cr v_R \cr} \right) , \quad
   \langle \Phi \rangle = \left(
                \matrix{v & 0 & 0 & 0  \cr
                        0 & v & 0 & 0  \cr
                        0 & 0 & v & 0  \cr
                        0 & 0 & 0 & v' \cr} \right) .
 \end{equation}
Electroweak doublets of Higgs fields
 \begin{equation}
   \chi_L \sim (1, 1, 2, 1; -1), \quad \chi_R \sim (1, 1, 1, 2; -1)
 \end{equation}
with vacuum expectation values
 \begin{equation}
   \langle \chi_L \rangle = \left( \matrix{w_L \cr 0 \cr} \right) , \quad
   \langle \chi_R \rangle = \left( \matrix{w_R \cr 0 \cr} \right)
 \end{equation}
are necessary to break the chiral symmetry and the Weinberg-Salam symmetry
in electroweak interaction. The vacuum expectation values are assumed to be
real and to satisfy the hierarchy
\begin{equation}
  v_R,\,v_L \gg v,\,v' > w_R > w_L,\ \ (v^2,\,v'{}^2 \gg w_R^2 \gg w_L^2).
\end{equation}

As pointed out in the introduction, it is necessary to specify the base of gauge
interaction eigenmodes of fundamental fermions in order to state the hypothesis
of the universal strength for Yukawa couplings in the SDUSM without ambiguity.
Here, we take the base in which a {\it bare-mass} matrix of the colourless
neutral fields $N'_{ih}$ is diagonal, and scalar interactions between
electroweak singlets $F_{iL}$ and $F_{iR}$ are also diagonal. Our hypothesis
on the USY in the SDUSM then is that the coupling constants for the Yukawa
interactions have generation dependence described by phase factors.
We express the {\it Yukawa coupling constants} for
scalar interactions between $\psi_{iL}$ and $F_{jR}$ by $Y_fe^{i\phi_{ij}^f}$
and those between $F_{iL}$ and $N'_{jR}$ by $Y'e^{i\varphi_{ij}^N}$, where
$\phi_{ij}^f$ and $\varphi_{ij}^N$ are real, $f = (u, d)$ and $F = (U, D)$.
Without loss of generality, the interactions of the bi-quartet Higgs field $\Phi$
with the electroweak singlet fields $F_{iL}$ and $F_{iR}$ are set to be described
by a coupling constant $Y_F$ in a generation-independent way since phase factors
can be eliminated by adjusting phases of the singlet fields.\par

The most general form of the Lagrangian density ${\cal L}_Y$ for the bare masses
and scalar interactions of fermions, which satisfies the USY and is invariant
under the left-right symmetric gauge group $G$, is written as
 \begin{equation}
 \begin{array}{lll}
   {\cal L}_Y & = & \ \sum_{i,j} \left\{
                 Y_u e^{i\phi_{ij}^u}
                 \left(
                {\bar \psi}_{iL} \chi_L U_{jR} + {\bar \psi}_{iR} \chi_R U_{jL}
                 \right)\right.\\
              \noalign{\vskip 0.3cm}
              & &\quad \ \ + \left. \,Y_d  \,e^{i\phi_{ij}^d}
                  \left(
                 {\bar \psi}_{iL} {\tilde \chi}_L D_{jR}
                 + {\bar \psi}_{iR} {\tilde \chi}_R D_{jL}
                  \right)
                           \right\} \\
              \noalign{\vskip 0.3cm}
              & & +  \sum_{i,j} Y' e^{i\varphi_{ij}^N}
                  \left({\bar N}'_{iL} \chi_{cL}^\dagger U_{jR} 
                  + {\bar N}'_{iR} \chi_{cR}^\dagger U_{jL} \right) \\
              \noalign{\vskip 0.3cm}
              & & - \sum_i \left(
                    Y_U{\bar U}_{iR} \Phi U_{iL} + Y_D{\bar D}_{iR} \Phi D_{iL}
                           \right) \\
              \noalign{\vskip 0.3cm}
              & & + \sum_i m_i^N {\bar N'}_{iL} N_{iR}' + {\rm h.c.},
 \end{array}
 \end{equation}
where $\tilde{\chi} = i \sigma_2 \chi^\ast$ and
$m_i^N (m_1^N\!<\!m_2^N\!<\!m_3^N)$ are the bare Dirac masses
of the colourless neutral fields $N'_{ih}$.
Henceforth the strengths of the Yukawa coupling constants $Y_f$, $Y'$ and $Y_F$
are considered to have nearly the same order of magnitude. \par

\section{Seesaw mass matrices with pure phase submatrices}

Spontaneous breakdowns of the underlying symmetry $G$ induce $6 \times 6$
seesaw mass matrices for the charged-fermion sectors and a $9 \times 9$ seesaw
mass matrix for the neutral-fermion sector. For general representations of
mass matrices, it is convenient to use the following bases in the generation
space as
\begin{equation}
  \begin{array}{lll}
   f = \left( \matrix{f_1 \cr f_2 \cr f_3} \right),\ &
   F = \left( \matrix{F_1 \cr F_2 \cr F_3} \right),\ &
   F' = \left( \matrix{F'_1 \cr F'_2 \cr F'_3} \right),
  \end{array}
\end{equation}
where the components $f_i$, $F_i$ and $F'_i$ are the multiplets $\psi_{ih}$,
$(U_{ih}, D_{ih})$ and $N'_{ih}$, respectively. The mass matrices are expressed
in the seesaw block-matrix forms
 \begin{equation}
   \left( \matrix{{\bar f}_L & {\bar F}_L \cr} \right)
   \left( \matrix{  0   & M_L^f \cr
                  M_R^f &  M^F  \cr} \right)
   \left( \matrix{f_R \cr F_R \cr} \right) + {\rm h.c.},
 \label{6x6}
 \end{equation}
for the charged-fermion sectors in the $(f, F)$ base, and
 \begin{equation}
   \left( \matrix{{\bar f}_L & {\bar F}_L & {\bar F}_L' \cr} \right)
   \left( \matrix{  0   &  M_L^f   &    0      \cr
                  M_R^f &   M^F    & M_R^{F'}  \cr
                    0   & M_L^{F'} &  M^{N'}   \cr} \right)
   \left( \matrix{f_R \cr F_R \cr F_R' \cr} \right) + {\rm h.c.},
 \label{9x9}
 \end{equation}
for the neutral-fermion sector in the $(f, F, F') = (\nu, N, N')$ base.\par

The $3\times 3$ submatrices $M_L^f$ and $M_R^f$ are given by
 \begin{equation}
   M_L^f = Y_f w_L M_\phi^f, \quad M_R^f = Y_f w_R M_\phi^{f\dagger}
 \label{MLMR}
 \end{equation}
in terms of the pure phase matrices
 \begin{equation}
   M_\phi^f \equiv
   \left( \matrix{e^{i\phi_{11}^f} & e^{i\phi_{12}^f} & e^{i\phi_{13}^f} \cr
                  e^{i\phi_{21}^f} & e^{i\phi_{22}^f} & e^{i\phi_{23}^f} \cr
                  e^{i\phi_{31}^f} & e^{i\phi_{32}^f} & e^{i\phi_{33}^f} \cr
                 } \right) .
 \label{phi}
 \end{equation}
The submatrix $M^F$ has the structure
 \begin{equation}
   M^F = - Y_F v E, \quad M^F = - Y_F v' E
 \end{equation}
for the quark and lepton sector, respectively, where $E$ is the $3\times3$ unit
matrix. As in Eq.\,(\ref{MLMR}), the submatrices $M_L^{F'}$ and $M_R^{F'}$ are
expressed as
 \begin{equation}
   M_L^{F'} = Y'v_L M_\varphi^{N}, \quad M_R^{F'} = Y'v_R M_\varphi^{N\dagger}
 \label{twist}
 \end{equation}
in terms of the pure phase matrix
 \begin{equation}
   M_\varphi^{N} \equiv \left( \matrix{e^{i\varphi_{11}^N} & e^{i\varphi_{12}^N}
                                      & e^{i\varphi_{13}^N} \cr
                           e^{i\varphi_{21}^N} & e^{i\varphi_{22}^N}
                                      & e^{i\varphi_{23}^N} \cr
                           e^{i\varphi_{31}^N} & e^{i\varphi_{32}^N}
                                      & e^{i\varphi_{33}^N} \cr
                 } \right).
 \label{varphi}
 \end{equation}
$M^{N'}$ is the diagonal matrix
 \begin{equation}
   M^{N'} = \left( \matrix{m_1^N & 0     & 0     \cr
                           0     & m_2^N & 0     \cr
                           0     & 0     & m_3^N \cr}
            \right).
 \end{equation}

\section{Effective mass matrices}

On the assumption that $Y_F^2v^2, Y_F^2v'{}^2 \gg Y_f^2w_R^2 \gg Y_f^2w_L^2$,
the seesaw mass matrices in Eqs.\,(\ref{6x6}) and (\ref{9x9}) are
block-diagonalized by the bi-unitary transformations. The single universal
seesaw mechanism results in the effective mass matrices
 \begin{equation}
   M_{\rm eff}^f = - M_L^f (M^F)^{-1} M_R^f
 \label{effmasscharged}
 \end{equation}
for ordinary low-lying charged fermions. Substituting the submatrices,
the effective mass matrices of the charged fermions are written in the form
 \begin{equation}
   M_{\rm eff}^f = M_f \tO_f,
 \end{equation} 
where
 \begin{equation}
  \tO_f = {1 \over 3}
   \left( \matrix{1 & a_3^f \, e^{ i\delta_{12}^f} &
                      a_2^f \, e^{-i\delta_{31}^f}     \cr
                  \noalign{\vskip 0.3cm}
                  a_3^f \, e^{-i\delta_{12}^f} & 1 &
                      a_1^f \, e^{ i\delta_{23}^f}     \cr
                  \noalign{\vskip 0.3cm}
                  a_2^f \, e^{ i\delta_{31}^f}     &
                      a_1^f \, e^{-i\delta_{23}^f} & 1  }
   \right).
 \label{quasidemocratic}
 \end{equation}
Here, $a_i^f$ and $\delta_{ij}^f$ are real parameters, which are expressed
in terms of the original phases $\phi_{ij}^f$ in Eq.\,(\ref{phi}), as follows:
 \begin{equation}
  a_i^f \, e^{ i\delta_{jk}^f} = {1 \over 3}\sum_l e^{i(\phi_{jl}^f-\phi_{kl}^f)}
 \end{equation}
where $(i, j, k) = {\rm cyclic}(1, 2, 3)$.
The mass scales $M_f\ (f = u, d, l)$ are fixed by
 \begin{equation}
  M_u = 9\displaystyle{{Y_u^2 \over Y_U}{w_Lw_R \over v}},\quad
  M_d = 9\displaystyle{{Y_d^2 \over Y_D}{w_Lw_R \over v}},\quad
  M_l = 9\displaystyle{{Y_d^2 \over Y_D}{w_Lw_R \over v'}.}
  \label{udlscale}
 \end{equation}

The masses of ordinary charged fermions $m_j^f$ and the eigenvalues $\to_j^f$
of the Hermitian matrix $\tO_f$ are related as follows:
 \begin{equation}
  \begin{array}{l}
   m_j^f = M_f \to_j^f,  \quad \sum_j \to_j^f = 1.
  \end{array}
 \label{massomega}
 \end{equation}
From Eqs.\,(\ref{udlscale}) and (\ref{massomega}), we find
 \begin{equation}
   {m_u + m_c + m_t \over m_d + m_s + m_b} =
   {Y_u^2 \over Y_d^2}{Y_D \over Y_U}
 \label{uoverd}
 \end{equation}
for the masses of up- and down-quark sectors. Owing to the large symmetry of
the group $G$, the quasi-democratic matrix is common to the down-quark and
charged-lepton sectors, i.e. $\tO_d = \tO_l$. Therefore, their mass
spectra have the same infrastructure. From Eqs.\,(\ref{udlscale})
and (\ref{massomega}), we obtain the relations
 \begin{equation}
   {m_e \over m_d} = {m_\mu \over m_s} = {m_\tau \over m_b} = {v \over v'}
 \label{loverd}
 \end{equation}
for the masses of the charged-lepton and down-quark sectors. The quasi-democratic
mass matrices $M_{\rm eff}^f$ in Eq.\,(\ref{quasidemocratic}) have been derived
and analysed in the previous articles~\cite{soga3}. We can inherit here all
the results of those analyses. \par

Provided that
$Y'{}^2v_L^2, Y'{}^2v_R^2 \gg Y_U^2v'{}^2, m_i^2 \gg Y_f^2w_R^2 \gg Y_f^2w_L^2$,
the double universal seesaw mechanism~\cite{soga1} leads to the effective mass
matrix
 \begin{equation}
  \begin{array}{lll}
   M_{\rm eff}^\nu & = & - M_L^\nu \left(M^N -  M_R^{N'}
                     \left( M^{N'} \right)^{-1} M_L^{N'}\right)^{-1} M_R^\nu \\
                     \noalign{\vskip 0.1cm}
                   & \approx & M_L^\nu  \left(M_L^{N'} \right)^{-1} M^{N'}
                               \left(M_R^{N'}\right)^{-1} M_R^\nu \\
                     \noalign{\vskip 0.2cm}
                   & \approx & \displaystyle{\left({Y_u \over Y'}\right)^2
                     {w_Lw_R \over v_Lv_R}}
                     [M_\phi^u\left(M_\varphi^N\right)^{-1}] M^{N'}
                     [M_\phi^u\left(M_\varphi^N\right)^{-1}]^\dagger
  \label{neutrionmassmatrix}
  \end{array}
 \end{equation} 
for the neutrino sector. While the quasi-democratic mass matrix
$M_{\rm eff}^f$ for the charged fermion sector is made out of
9 phases $\phi_{ij}^f$ of the matrix in Eq.\,(\ref{phi}),
the effective mass matrix $M_{\rm eff}^\nu$ for the neutrino sector includes 18
phases $\phi_{ij}^u$ and $\varphi_{ij}^N$ of the matrices in Eqs.\,(\ref{phi})
and (\ref{varphi}). Without a skillful reduction of these unknown phases,
it is impossible to determine a concrete form of $M_{\rm eff}^\nu$.
Here, resorting to the principle of simplicity,
we postulate an additional ansatz that the pure phase matrices
$M_\phi^u$ and $M_\varphi^N$ are identical, i.e.
$\phi_{ij}^u = \varphi_{ij}^{N},\ {\rm mod}(2\pi)$.
This ansatz enables us to eliminate all the unknown phases and to simplify
the effective mass matrix $M_{\rm eff}^\nu$ into the diagonal form
 \begin{equation}
    M_{\rm eff}^\nu = S_\nu M^{N'},\quad
    S_\nu = \left({Y_u \over Y'}\right)^2{w_Lw_R \over v_Lv_R}.
 \end{equation}
Accordingly we obtain the simple mass formula for the neutrions as follows:
 \begin{equation}
  m_i^\nu = S_\nu m_i^N.
 \end{equation}

\section{Vacuum neutrino oscillation}

The weak mixing matrix $V$ in the leptonic sector is determined by deriving a
diagonalization matrix for the effective mass matrix $M_{\rm eff}^l$ of the
charged lepton sector since the neutrino mass matrix is already given in the
diagonal form. In order to obtain the diagonalization matrix, which is calculable
in terms of the charged-lepton masses exclusively, {\it CP}-violation effects are
neglected and a simplification $a_1^l = a_2^l$ is made in the matrix $\tO_l$. 
Following the method used in the previous paper~\cite{soga3}, we introduce the
following parametrization as
\begin{equation}
  a_1^l = a_2^l = {3 \over 2\sqrt{2}}\rho_l\sin\theta_l, \quad
  a_3^l = 3\rho_l\cos\theta_l.
\end{equation}
Then the eigenvalue problem for the matrix
 \begin{equation}
   M_{\rm eff}^l = M_l \tO_l \approx M_l \Omega_l
 \end{equation}
with
 \begin{equation}
  \Omega_l = {1 \over 3}
    \left(
     \begin{array}{ccc}
       1                    &
       3\rho_l\cos\theta_l  &  {3 \over 2\sqrt{2}}\rho_l\sin\theta_l \\
       \noalign{\vskip 0.3cm}
       3\rho_l\cos\theta_l  &
       1                    &  {3 \over 2\sqrt{2}}\rho_l\sin\theta_l \\
       \noalign{\vskip 0.3cm}
      {3 \over 2\sqrt{2}}\rho_l\sin\theta_l &
      {3 \over 2\sqrt{2}}\rho_l\sin\theta_l &  1  \\
     \end{array}
   \right)
   \label{simplified}
 \end{equation}
turns out to be solvable in a simple analytical form. With the orthogonal matrix
 \begin{equation}
   U_l = {1 \over \sqrt{2}}
    \left(
     \begin{array}{ccc}
       1                  &  -1                  &   0                         \\
       \noalign{\vskip 0.2cm}
       \sin\half\theta_l  &  \sin\half\theta_l  &  -\sqrt{2}\cos\half\theta_l  \\
       \noalign{\vskip 0.3cm}
       \cos\half\theta_l  &  \cos\half\theta_l  &  \ \sqrt{2}\sin\half\theta_l \\
     \end{array}
   \right),
   \label{ortho}
 \end{equation}
the quasi-democratic matrix $\Omega_l$ is diagonalized as
 \begin{equation}
   U_l\Omega_lU_l^\dagger 
   = {\rm diag}\left( \omega_1^l,\, \omega_2^l,\, \omega_3^l \right),
 \end{equation}
where
 \begin{equation}
  \omega_1^l = \i3 - \rho_l\cos\theta_l,\ 
  \omega_2^l = \i3 + \half\rho_l(-1 + \cos\theta_l),\ 
  \omega_3^l = \i3 + \half\rho_l( 1 + \cos\theta_l).
 \end{equation}
The parameters $\rho_l$ and $\cos\theta_l$ are expressed in terms of
the charged-lepton masses by
 \begin{equation}
  \rho_l = 1 - { 2m_\mu + m_e \over m_\tau + m_\mu + m_e},\quad
  \cos\theta_l = \i3\left(1 + 2{m_\mu - m_e \over m_\tau - m_\mu}\right).
 \label{costheta}
 \end{equation}
Since the diagonalization matrix $\Omega_l$ depends only on the parameter
$\theta_l$, the weak mixing matrix $V = U_l$ is calculable in terms of
the charged-lepton masses in this approximation.
Note that the matrix $\Omega_l$ takes the democratic form and its eigenvalues
results in $(\omega_1^l,\, \omega_2^l,\, \omega_3^l) = (0,\,0,\,1)$ at the
limit $\rho_l = 1$ and $\cos\theta_l = \i3$. 

The experimental data of neutrino anomalies are usually presented in terms of
two flavour-mixing parameters $\Delta{m_{ij}^2} = |m_i^2 -m_j^2|$ and
$\sin^22\theta_{ij}$. The most natural solution of the atmospheric neutrino
anomaly is considered to be given by the $\nu_\mu \leftrightarrow \nu_\tau$
oscillation with the following ranges of parameters~\cite{superk,atmospheric}
 \begin{equation}
  \Delta{m_{\rm atm}^2} \approx (0.5 - 6)\times 10^{-3}{\rm eV}^2,\quad
  \sin^22\theta_{\rm atm} > 0.82
  \label{atmosphericdata}
 \end{equation}
If we apply the mechanism of neutrino oscillation to interpret also the
phenomenon of the solar-neutrino deficit, the most probable candidate is the
$\nu_e \leftrightarrow \nu_\mu$ oscillation, with the following ranges of
parameters~\cite{solar,bahcall}:
 \begin{equation}
  \Delta{m_{\rm sol}^2} \approx (0.6 - 1.1)\times 10^{-10}{\rm eV}^2,\quad
  \sin^22\theta_{\rm sol} \approx 0.7 - 1.
  \label{solardata}
 \end{equation}
In these interpretations where the parameters $\Delta{m_{ij}^2}$ satisfy
$\Delta{m_{21}^2} \ll \Delta{m_{32}^2} \approx \Delta{m_{31}^2}$,
the angles of two-flavour mixings can be approximated by the matrix elements of
the three-flavour mixing matrix $V = U_l$ as 
 \begin{equation}
   \sin^2 2\theta_{\rm atm} = 4 |V_{23}|^2 (|V_{21}|^2 + |V_{22}|^2)
 \end{equation}
and
 \begin{equation}
   \sin^22\theta_{\rm sol} = 4 |V_{11}|^2 |V_{12}|^2
 \end{equation}
in good approximation. \par

Using Eqs.\,(\ref{ortho}) and (\ref{costheta}), we obtain
 \begin{equation}
   \sin^2 2\theta_{\rm atm}
    = \sin^2\theta_l
    = {4 \over 9}\left(1 - {m_\mu - m_e \over m_\tau - m_\mu}\right)
                 \left(2 + {m_\mu - m_e \over m_\tau - m_\mu}\right)
 \label{atmospherictheta}
 \end{equation}
and
 \begin{equation}
   \sin^22\theta_{\rm sol} = 1.
 \label{solartheta}
 \end{equation}
These values representing the large mixings are consistent with the experimental
data in Eqs.\,(\ref{atmosphericdata}) and (\ref{solardata}). The latter is
identical to the result obtained at the democratic limit by Fritzsch and
Xing~\cite{fritzschxing} and the former reduces to the relation
$\sin^22\theta_{\rm atm} = 8/9$ discovered by them at the same limit.
\par

As for the neutrino masses, two interpretations are often advocated, i.e.
$(i)$ the almost degenerate spectrum with
$m_1^\nu \approx m_2^\nu \approx m_3^\nu$ and $(ii)$ the hierarchical spectrum
with $m_1^\nu \ll m_2^\nu \ll m_3^\nu$. Both interpretations are applicable to
the present model. Since $m_i^\nu = S_\nu m_i^N$, these characteristics of the
neutrino mass spectrum are translated into those of the exotic neutral fields.
In the former case, it is possible to identify the hot component of dark
matter in the Universe with the neutrinos that have $m_i^\nu \approx
2.5$ eV~\cite{darkmatternu}. \par

\section{Discussion}

For the up-quark sector the seesaw approximation is usually considered to be
disqualified, since the magnitude of the Yukawa coupling constant has to be
much larger than 1 for the top-quark mass $m_t$ to be close to the electroweak
scale $w_L$. In our theory the numerical factor 9, which appears through
the product of $M_\phi^fM_\phi^{f\dagger}$ in Eq.\,(\ref{effmasscharged}), acts
to improve the situation considerably. Equation (\ref{udlscale}) and the relation
$M_u \approx m_t \approx w_L$ lead to the estimate $Y_Uv \approx 9Y_u^2w_R$.
This estimate tells us that the condition for the seesaw approximation
$Y_U^2v^2 \gg Y_u^2w_R^2$ holds if the criterion $81Y_u^2 \gg 1$
is satisfied. Therefore, for the Yukawa coupling constant $Y_u$ of the order,
say, of around $1/2$, the seesaw approximation turns out to be applicable to
the up-quark sector.\par 

Departures of the mass scales of the down-quark and charged-lepton
sectors from the electroweak scale $w_L$ must be explained, respectively,
by the seesaw factors $9{Y_d^2 w_R / Y_Dv}$ and $9{Y_d^2 w_R / Y_Dv'}$.
The scale difference of up- and down-quark sectors has to be ascribed to
the difference between the strengths of the Yukawa coupling constants $Y_f$
and $Y_F$. Namely those constants must be tuned so as to satisfy the relation
$Y_u^2Y_D/Y_d^2Y_U \simeq  m_t/m_b$. This tuning seems to be executable without
changing the orders of the constants $Y_f$ and $Y_F$. In the present stage of
our model, the down-quark and charged-lepton sectors have similar mass matrices
and mass spectra.  It is necessary to find a mechanism of $SU(4)_h$ symmetry
breakings to generate fine variations in these two sectors. To explain
the scale difference of the spectra, the vacuum expectation values of
the bi-quartet $\Phi$ must satisfy the relation $v/v' \simeq m_\tau/m_b$. \par

The masses of the neutrinos and the exotic neutral fields are proportional to
each other. To explain the strong suppression of the neutrino mass scale, vacuum
expectation values of $\chi_{ch}$ must be assumed to be sufficiently larger
than those of $\chi_h$. Note that, owing to the result of the CERN precision
measurement~\cite{lepewwg}, the bare masses of the neutral exotic fields must
be larger than the half-value of the $Z$ boson mass, i.e. $m_i^N > \half m_Z$.
Therefore, if we adopt the almost degenerate mass spectrum $(i)$ for the
neutrinos and identify the hot dark matter with them, the scale factor
$S_\nu \approx {w_Lw_R/v_Lv_R}$ is subject to the restriction
$S_\nu < 2 m_i^\nu/m_Z \approx 5\times 10^{-11}$.
In the case of the hierarchical mass spectrum $(ii)$, the scale factor must
satisfy the more stringent relation
$S_\nu < 2 m_1^\nu/m_Z \ll 2 m_2^\nu/m_Z \approx 10^{-16}$.
Therefore the breakdowns of the $SU(4)_h$ colour symmetries have to occur at
much higher energy scales in the hierarchical case $(ii)$ than in the degenerate
case $(i)$. Here it should be mentioned that, in our model with the USM, no
Majorana mass is assumed to exist. All masses are of the Dirac type. Therefore
the observed bound of the neutrinoless $\beta\beta$-decay~\cite{betabeta} does
not impose any restriction on our theory. \par

In this model the hypothesis of the USY is stated in the simplest fermion base,
where the neutral exotic fields have the diagonal mass matrix and the
electroweak singlets of colour quartets interact in the generation diagonal
manner. It is possible to choose other fermion bases. If an off-diagonal mass
matrix is adapted for the neutral exotic fields, it affects the weak mixing
matrix and accordingly the estimates in Eqs.\,(\ref{atmospherictheta})
and (\ref{solartheta}). The additional ansatz $M_\phi^u = M_\varphi^N$ plays
the decisive role to render the simple mass matrix for the neutrinos by totally
eliminating the contributions from unknown submatrices in this model. 
Its physical and geometrical meanings must be investigated. \par

\bigskip
\noindent
{\it{\bf Acknowledgements\ }}\ One of the authors, I.~S.~Sogami, would like to
express his sincere thanks to H.~Fritzsch and Z.~Z.~Xing for useful comments
and discussions.

\end{document}